\documentclass[aps,prb,twocolumn,superscriptaddress,showpacs,amsmath,amssymb]{revtex4-1}
\usepackage{amsmath}
\usepackage{amsfonts}
\usepackage{amssymb}
\usepackage{graphicx}
\usepackage{color}

\usepackage{bm}
\usepackage{braket}
\usepackage{gensymb}

\usepackage{ulem}

\begin{document}

\title{Quantum valence bond ice theory for proton-driven quantum spin-dipole liquids}

\author{Masahiko G. Yamada}
\email[]{myamada@mp.es.osaka-u.ac.jp}
\affiliation{Department of Materials Engineering Science, Osaka University, Toyonaka 560-8531, Japan}
\affiliation{Institute for Solid State Physics, University of Tokyo, Kashiwa 277-8581, Japan}
\affiliation{Department of Physics, Massachusetts Institute of Technology, Cambridge, Massachusetts 02139, USA}
\author{Yasuhiro Tada}
\email[]{tada@issp.u-tokyo.ac.jp}
\affiliation{Institute for Solid State Physics, University of Tokyo, Kashiwa 277-8581, Japan}

\date{\today}

\begin{abstract}
We present a theory of a hybrid quantum liquid state,
\textit{quantum spin-dipole liquid} (QSDL), in a hydrogen-bonded electron system,
by combining a quantum proton ice and Anderson's resonating valence bond spin liquid theory,
motivated by the recent experimental discovery of a quantum spin liquid with proton fluctuations
in $\kappa$-H$_3$(Cat-EDT-TTF)$_2$ (a.k.a. H-Cat). In our theory,
an electron spin liquid and a proton dipole liquid are realized simultaneously in the ground state
called \textit{quantum valence bond ice}.
In this state, neither of them can be established independently of the other.
Analytical and numerical calculations
reveal that this state has a large entanglement entropy between spins and dipoles,
which is far beyond the (crude) Born-Oppenheimer approximation.
We also examine the stability of QSDL with respect to perturbations and discuss
implications for experiments in H-Cat and its deuterated analog D-Cat.
\\ \\
PhySH: Frustrated magnetism, Spin ice, Spin liquid, Quantum spin liquid, Organic compounds
\end{abstract}

\maketitle

\section{Introduction}
Hydrogen bonds occupy a unique position in condensed matter physics,
where we can expect a huge quantum paraelectricity due to its lightness.
Most hydrogen-bonded systems so far were treated within the (crude) Born-Oppenheimer approximation,
where we ignore quantum entanglement between atomic motion and electronic states.
However, recently it was recognized that in many hydrogen-bonded materials quantum nature
of protons plays an essential role and even affects
basic properties of the system~\cite{Morrone2009,Li2011,QNE2018,Reiter2002,Bove2009ice,Morimoto1991}.
Indeed, protons can tunnel
quantum mechanically between two stable positions of the O--H$\cdots$O bonding
and furthermore the hydrogen bonds are subject to the ice rule due to the frustration
in a wide range of systems, including
ferroelectric KH$_2$PO$_4$~\cite{Reiter2002,Koval2005}, hexagonal water ice~\cite{Bove2009ice,Benton2016},
and molecular crystals~\cite{Morimoto1991,Ishizuka2011,Chern2014}.

Recently, a new member of hydrogen-bonded materials,
purely organic \mbox{$\kappa$-H$_3$(Cat-EDT-TTF)$_2$} (called H-Cat),
was synthesized~\cite{Isono2013,Isono2014,Ueda2014,Ueda2015,Yamashita2017,Shimozawa2017qsdl,Yamamoto2016,Tsumuraya2015,Naka2018}.
H-Cat has a quasi-two-dimensional (2D) triangular lattice structure of dimers with moderate interlayer coupling,
and each layer is connected by
hydrogen bonds which also form a triangular lattice.
Experimentally, protons are unfrozen down to the lowest temperature
and the observed quantum paraelectricity can be attributed to proton tunneling~\cite{Shimozawa2017qsdl}.
In stark contrast to the other hydrogen-bonded materials, H-Cat is
shown to be a dimer Mott insulator, where we can expect strong electron correlation.
Moreover, in H-Cat, spins delocalized over the $\pi$-conjugated orbital
are also fluctuating without any magnetic order
below a temperature scale close to that of the quantum proton motion.
This implies that a quantum liquid state of unknown type is realized in H-Cat,
where both protons and electrons are quantum mechanically disordered simultaneously
and are strongly entangled with each other, which makes this material far beyond the (crude) Born-Oppenheimer approximation.
On the other hand, the deuterated counterpart D-Cat exhibits a charge density wave (CDW) state
below 185~K,
where protons and electrons together show a coupled charge order,
indicating strong Coulomb interaction between them~\cite{Isono2013,Ueda2014,Shimozawa2017qsdl}.
Thus,
such a large isotope effect strongly suggests that the proton tunneling is responsible
for the possible hybrid quantum liquid state in H-Cat.
However, it is unclear why proton and spin fluctuations can coexist in a quantum
liquid state.

The triangular lattice of H-Cat is a prototypical example of frustrated lattices and
such a frustrated lattice can impose constraints on proton and spin configurations,
leading to macroscopic degeneracy at a classical level.
Quantum fluctuations could potentially realize a quantum liquid state out of the degenerate classical states
as originally proposed in Anderson's theory
of the resonating valence bond (RVB) state~\cite{Anderson1973RVB,Balents2010qsl,Savary2016}.
The basic mechanism of realizing a quantum liquid was
later employed in the theory of quantum spin/water ice where the RVB state is exactly realized
by tuning a system into the exactly solvable Rokhsar-Kivelson (RK) point~\cite{Balents2010qsl,Savary2016,Moessner2003,Hermele2004,CastroNeto2006,Shannon2012}.
This observation would imply that both the electron spin liquid and the proton dipole liquid
can be treated in a unified manner to describe a spin-dipole coupled quantum liquid.

In this paper, motivated by the experimental discovery
of H-Cat,
we present a theory of a hybrid quantum liquid state,
\textit{quantum spin-dipole liquid} (QSDL),
in hydrogen-bonded electron systems
by combining the concepts of Anderson's RVB state of electron spins and a quantum ice of protons.
Although a hybrid quantum spin liquid itself is discussed in the literature~\cite{Nakatsuji2012qsol,Corboz2012qsol,Yamada2018su4,Kitaev2006,Kitagawa2018qsl,Smerald2018lattice,sl_RMP2017},
including spin-charge coupled purely electron systems without hydrogen bonds~\cite{Hotta2010qsdl,Naka2010qsdl,Naka2016qsdl,Yakushi2015,Sedlmeier2012}, 
its entangled nature has never been discussed in detail.
In our theory, the RVB state is extended to describe the entanglement between spins
and dipoles.
We give a simple concrete description of the QSDL based on a minimal model which
incorporates three key features of the QSDL, i.e. macroscopic classical degeneracy,
quantum proton tunneling, and electron correlations.
The resulting QSDL is named \textit{quantum valence bond ice}, and the property of this
state will be discussed in detail from analytical and numerical perspectives.
Though this unified theory does not explain everything observed in H-Cat, especially
the absence of a spin gap~\cite{Shimozawa2017qsdl}, we can understand some basic difference between
H-Cat and D-Cat based on a universal nature of a quantum liquid.  Thus, while the model
itself is idealized to some extent, it can give a clear physical picture of the state
of the state observed in H-Cat.

\section{Model}
\begin{figure}[t]
\includegraphics[width=8.6cm]{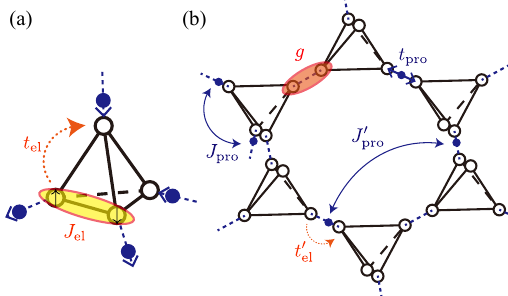}
\caption{(a) Tetrahedron of electron sites. (b) Diamond lattice network of proton bonds.
Blue dashed lines represent proton sites and white circles represent electron sites.
In the following, yellow ellipses represent valence bonds of holes.
Blue solid lines represent proton interactions and orange dotted lines represent
electron/hole hoppings. A red circle represent the electron-proton interaction.
For later purposes, an additional proton-proton interaction $J'_\textrm{pro}$ and
an intertetrahedron hole hopping $t'_\textrm{el}$ are also shown.}
\label{fig:lattice}
\end{figure}

Since the primary purpose of the present study is to present a firm basis for hybrid spin liquids, it is desirable to introduce an appropriate model which is inspired by the candidate material H/D-Cat.
First, we focus on the electron sector of H/D-Cat where there are two inequivalent dimers (four sites) in the unit cell,
and thus
consider an idealized model where 
tetramers are introduced as a fundamental unit of the system.
The tetramer would be regarded as a pair of dimers, and therefore the model can describe not only QSDL
relevant to an H-bonded system but also a CDW state corresponding to a D-bonded system on an equal footing,
as seen in H/D-Cat. 
Our Hamiltonian contains both holes and protons, and is defined by
\begin{align}
		H &= H_\textrm{el}+H_\textrm{pro}+H_\textrm{el--pro}, \label{eq:H} \\
		H_\textrm{el} &= t_\textrm{el} \sum_{\boxtimes}\sum_{i,j\in \boxtimes} c_{is}^\dagger c_{js} + J_\textrm{el} \sum_{\boxtimes}\sum_{i,j\in \boxtimes} \bm{S}_i \cdot \bm{S}_j, 
\label{eq:tJ}\\
		H_\textrm{pro} &= t_\textrm{pro} \sum_{\langle ij\rangle} \sigma_{ij}^x + J_\textrm{pro} \sum_{\langle\langle ij \rangle,\langle kl \rangle\rangle} \sigma_{ij}^z \sigma_{kl}^z, \\
		H_\textrm{el--pro} &= g \sum_{\langle ij \rangle} (n_j-n_i)\sigma_{ij}^z.
\label{Ham}
\end{align}
The holes are described by the $t$--$J$ model where double occupancy is prohibited
by the strong onsite interaction~\footnote{Mathematically speaking, this condition is unnecessary for the following discussion}:
a real $t_\textrm{el}$ is a hole hopping parameter inside a tetrahedron denoted by $\boxtimes$, 
and $J_\textrm{el}>0$ is a spin-spin exchange interaction
between nearest-neighbor (NN) holes [see Fig.~\ref{fig:lattice}].
$c_{js}$ is an annihilation operator of a hole at the $j$th site with a spin $s,$
and $\bm{S}_j$ are spin-1/2 operators defined for this hole.
The summation over $s=\uparrow,\downarrow$ is implied.
$\bm{\sigma}_{ij}$ are Pauli matrices
defined on each hydrogen bond $\langle ij\rangle$ connecting the NN tetrahedra,
and two eigenstates with eigenvalues $\pm 1$ of
the $\sigma_{ij}^z$ operator correspond to the stable positions of a proton on the hydrogen bond $\langle ij\rangle$~\cite{Shimozawa2017qsdl,Naka2018}.
The tetrahedra sit on vertices of a diamond lattice and $\sigma^z_{ij}$ is defined such that
$\sigma^z_{ij}=+1$ for the sites $i(j)$ belonging to the A(B)-sublattice of the diamond lattice
corresponds to a proton position away from the A-sublattice site $i.$
A real $t_\textrm{pro}$ represents quantum tunneling between the two positions,
and $J_\textrm{pro}>0$ is a Coulomb repulsion between the protons
on a NN pair of bonds $\langle\langle ij \rangle,\langle kl \rangle\rangle.$
$g>0$ is a Coulomb repulsion between the hole and the proton.
We will later extend the model Eq.~\eqref{eq:H} and discuss effects of an additional proton-proton interaction
and intertetrahedron hole hopping.

To be concrete, we consider a range of model parameters by referring to H/D-Cat.
Since the dimer Mott insulator H-Cat is a one-hole-filled system~\cite{Isono2013,Ueda2014,Shimozawa2017qsdl,Tsumuraya2015,Naka2018},
we consider two-hole filling for each tetrahedron which is a counterpart with the same filling in our model.
Essentially the same argument can also be applied for a two-electron filling system.
While $H_\textrm{el}$ has a form
of the $t$--$J$ model, $H_\textrm{pro}$ is a transverse-field Ising model on the
frustrated pyrochlore lattice, and related models have been investigated as a model
for quantum spin liquids~\cite{Balents2010qsl,Savary2016,Moessner2003,Hermele2004,CastroNeto2006,Shannon2012}.
$g$ is supposed to be very large, \textit{i.e.}
$g \sim O(10^{3\textrm{--}4})$ K, because it comes from strong Coulomb repulsion.
$J_\textrm{pro}$ is also large enough, \textit{i.e.} $J_\textrm{pro} \sim O(100)$ K,
which is taken from the antiferroelectric transition temperature in D-Cat.
In addition, we suppose $|t_\textrm{pro}| \ll J_\textrm{pro}$ and $|t_\textrm{el}| \ll g$
by referring to the parameters in H-Cat where $t_\textrm{pro}\sim 1$--10 K and
$t_\textrm{el}\sim10$--100 K~\cite{Isono2013,Shimozawa2017qsdl,Tsumuraya2015,Naka2018,t_pro},
which allows a perturbation theory with respect to $t_\textrm{pro}.$
The magnetic coupling is assumed to be $J_\textrm{el} \sim10$--100 K taken from the value for H-Cat, and
one can focus only on the singlet sector of a tetrahedron
when the temperature is well below $J_\textrm{el}.$
This means that we consider a spin gapped state for simplicity, but our theory is
applicable even to the triplet sector just by replacing singlets by triplets,
and thus our theory can be extended to the case where a spin gap is zero.
In addition, a gauge mean-field theory is known to describe a fermionic excitation
from an ice state~\cite{Savary2012}, which would potentially describe the observed
exotic excitations~\cite{Shimozawa2017qsdl}.

\section{analysis of ground state}
\subsection{QSDL as a $\mathrm{U}(1)$ quantum liquid}
In this section, the ground state of the Hamiltonian Eq.~\eqref{eq:H} is discussed by 
mapping it to a quantum vertex model based on a perturbation theory.
In the case of $t_\textrm{pro}=0,$ the Hamiltonian
is exactly solvable
and the ground states are degenerate for all proton configurations
obeying the ice rule.
In the limit of $t_\textrm{el}/g\rightarrow0$, the proton configuration completely couples to the valence bonds of
hole pairs, and the position of the valence bond is fixed inside
each tetrahedron in the ground states. Since the valence bonds
also form an ice configuration, this state may be called \textit{classical valence bond ice}.
This basically holds true even for a nonzero $t_\textrm{el}$ where one can
diagonalize the electron Hamiltonian of each tetrahedron
in the valence bond basis and the ground state of a tetrahedron will be
a dressed valence bond state with a finite excitation gap $\sim g.$
Thus, every degenerate ground state of the total system at $t_\textrm{el}\neq 0$
is adiabatically connected from that at $t_\textrm{el}=0.$

\begin{figure}[t]
\includegraphics[width=8.6cm]{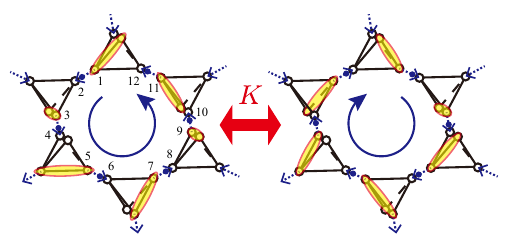}
\caption{The 6th-order perturbation process in $t_\textrm{pro}$ between the
flippable ice states.
Blue dashed arrows represent proton configuration and the corresponding valence bond configurations
are also shown.
All arrows obey the ice rule.
}
\label{fig:K}
\end{figure}

We now introduce a quantum tunneling of protons $t_\textrm{pro}\neq0,$
which will lift the macroscopic degeneracy of the classical ground states in the same way as in the quantum spin ice~\cite{Balents2010qsl,Savary2016,Moessner2003,Hermele2004,CastroNeto2006,Shannon2012}.
We can treat the effects of $t_\textrm{pro}$ when
$|t_\textrm{pro}|/J_\textrm{pro} \ll 1$ by using the degenerate perturbation theory.
Though essentially the same argument is true for a moderate $t_\textrm{el},$
we assume the condition $|t_\textrm{el}|/g\ll 1,$ which makes a physical picture clear.
Then, if a proton moves from its ground state position to break the ice rule,
a hole in the nearby tetrahedron will be pushed away from the proton
because of the repulsive interaction $g$ between them.
In this way, the proton motion is always accompanied by the hole hopping inside a tetrahedron and vice versa.
We note that such a charge-correlated hopping leads to effective
coupling between spins and protons despite the absence of an explicit spin-orbit interaction.
In this sense, spin fluctuations and proton fluctuations are tightly connected with each other, 
which will eventually lead to the QSDL state.
A nontrivial contribution appears from the 6th-order perturbation in $t_\textrm{pro}/J_\textrm{pro}$.
This process results in the following effective Hamiltonian in addition to a constant energy shift of the degenerate ground states,
\begin{align}
H_\textrm{eff}=-K\sum_\textrm{plaquette}
[\ket{\circlearrowleft}\bra{\circlearrowright}+\ket{\circlearrowright}\bra{\circlearrowleft}],
\label{eq:QIM}
\end{align}
where the flipping operator is given by the product of the hole and proton operators along a hexagonal plaquette loop as shown in Fig.~\ref{fig:K}:
$\ket{\circlearrowleft}\bra{\circlearrowright}=c^{\dagger}_{1}\sigma^+_{12}c_{2}c^{\dagger}_{3}\sigma^-_{34}c_{4}\cdots
c^{\dagger}_{11}\sigma^-_{11,12}c_{12}$ with $\sigma^{\pm}=(\sigma^x\pm i\sigma^y)/2$
in the lowest order approximation for $t_\textrm{el}/g$ [see Fig.~\ref{fig:K} for the definition of the site numbers].
The sum runs over the hexagonal plaquettes of the diamond lattice, and
the coefficient $K > 0$ is estimated to be $O(t_\textrm{pro}^6/J_\textrm{pro}^{5})$ with
a prefactor $O(t_\textrm{el}^6/g^6)$ when $|t_\textrm{el}|$ is small
(see Appendix \ref{App:A} for details).

The effective Hamiltonian Eq.~\eqref{eq:QIM} is formally equivalent to the
quantum ice model, which has been extensively studied in the context of the quantum spin
ice~\cite{Balents2010qsl,Savary2016,Moessner2003,Hermele2004,CastroNeto2006,Shannon2012}.
We stress, however, that the physical constituents of the low energy degrees of freedom
$\ket{\circlearrowleft},\,\ket{\circlearrowright}$ are very different from those in the quantum spin ice;
they are spin-dipole coupled degrees of freedom in the former, while solely spins in the latter.
The ground state of the effective Hamiltonian is known to be the quantum liquid state, which
is described by an emergent compact $\mathrm{U}(1)$ electromagnetic field which is deconfined in $3+1$ dimensions.
Thus, we have theoretically 
established that the quantum proton motion can indeed drive the system into a QSDL,
namely the quantum valence bond ice, and form a stable three-dimensional (3D) phase. 
It is stressed that both spins and protons are unfrozen simultaneously, forming a hybrid liquid state.
The coupled spins and dipoles behave as gapless ``photons'' of the emergent $\mathrm{U}(1)$ gauge theory,
while local proton excitations which break the ice rule with an energy gap $\sim J_\textrm{pro}$
are regarded as ``monopoles''~\cite{Balents2010qsl,Savary2016,Moessner2003,Hermele2004,CastroNeto2006,Shannon2012}.

One of the most characteristic features of the QSDL is a large entanglement between protons and holes.
The corresponding entanglement entropy defined by
$S_\textrm{EE}=-\textrm{Tr}_\textrm{el}\,\rho_\textrm{el}\log\rho_\textrm{el}$
with $\rho_\textrm{el}=\textrm{Tr}_\textrm{pro}\,\ket{\Psi}\bra{\Psi}$
for the ground state $\ket{\Psi}$ can provide essential information of its entanglement structure.
Now one can estimate that 
$S_\textrm{EE}\sim \log(\textrm{the number of flippable ice states})=
O(N_\textrm{tot})$, where $N_\textrm{tot}=N_\textrm{pro}+ N_\textrm{el}$
and $N_\textrm{pro/el}$ is the total number of proton bonds/electron sites.
This is easily understood for example based on the explicit calculation of $S_\textrm{EE}$
for the Rokhsar-Kivelson state~\cite{RK1988} (see Appendix \ref{App:B}).
Therefore, the QSDL has an entanglement entropy with a volume law
in contrast to the area law
for a real space bipartition~\cite{Stephan2009,EE_arealaw2010}.
Such a large entanglement between protons and holes is an intrinsic nature of the QSDL,
which is far beyond the (crude) Born-Oppenheimer approximation.

\begin{figure}[t]
\includegraphics[width=0.95\linewidth]{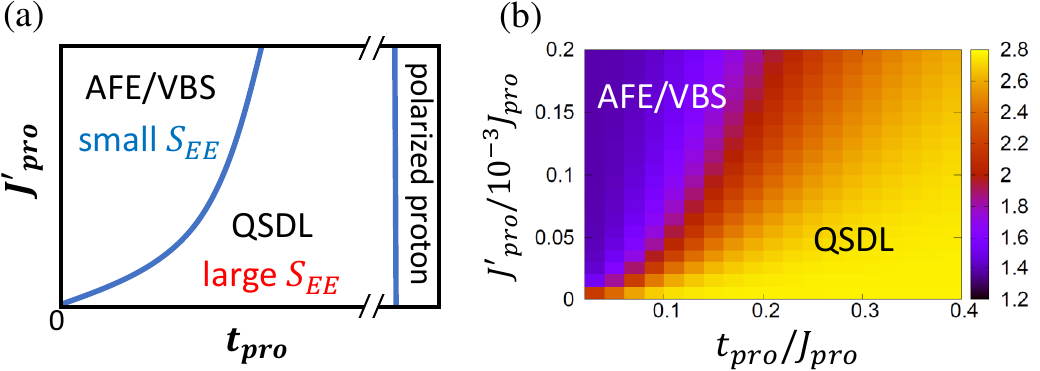}
\caption{(a) Schematic phase diagram for the 3D thermodynamic
system in the $t_\textrm{pro}$--$J'_\textrm{pro}$ plane.
The entanglement entropy $S_\textrm{EE}$ is small in the AFE/VBS phase,
while it is large exhibiting  the clear volume law $S_\textrm{EE}=O$(volume) in the QSDL phase.
``Polarized proton'' in the large $t_\textrm{pro}$ region means that protons are aligning in the direction of the ``transverse field'' $t_\textrm{pro}$.
(b) Entanglement entropy for the finite size 2D system Eq.~\eqref{eq:H_ED} calculated by the exact diagonalization.
The parameters used are $(t_\textrm{el},g,t'_\textrm{el})=(0.3,2.0,0.0)$
in units of $J_\textrm{pro}=1.0$.
}
\label{fig:EE}
\end{figure}

\subsection{Stability of QSDL}
Up to here, we have been based on the idealized model Eq.~\eqref{eq:H} where
a minimal number of terms are included to describe the QSDL.
However, we can consider additional terms in general, 
and considerations on such terms are important to discuss the relevance
of our model to real materials.
Here, we try to study the stability of the QSDL by introducing two types of perturbations into the model Eq.~\eqref{eq:H};
(i) an additional interaction between protons which favors a classical order by lifting the macroscopic
degeneracy of protons,
and (ii) an intertetrahedron hole hopping which may also lift the macroscopic degeneracy of electrons.
The following discussions will be supplemented with numerical calculations later.

First, we examine effects of additional proton interactions such as
$J'_\textrm{pro}\sum \sigma^z_{ij}\sigma^z_{kl}$
shown in Fig.~\ref{fig:lattice}(b),
and discuss implications for
the qualitative difference in $X$-Cat ($X=$ H, D).
The additional interaction lifts the macroscopic degeneracy of the classical ground states,
and an infinitesimal $J_\textrm{pro}'$ will lead to an ordered state when $t_\textrm{pro}=0$
by the order-by-disorder mechanism.
Indeed, classical ($t_\textrm{pro}=0$) spin ice models with additional interactions have been extensively studied,
where several ordered states are stabilized depending on types of the additional interactions~\cite{Melko2001,Chern_ice2008}.
In the present system, we consider $J_\textrm{pro}'$ interaction which favors an antiferroelectric (AFE) state
of protons,
but it competes with the quantum tunneling $t_\textrm{pro}$~\cite{Ishizuka2011,Savary2016,Savary2012}.
Based on these observations, one can draw a schematic phase diagram of the perturbed model with both
$J_\textrm{pro}'$ and $t_\textrm{pro}$ as shown in Fig.~\ref{fig:EE}(a),
where the QSDL with a large entanglement entropy remains stable for $J_\textrm{pro}'\ll K,$
while the AFE state with a small entanglement is favored for $J_\textrm{pro}'\gg K.$
Note that, because $g\gg |t_\textrm{el}|$, holes simultaneously
show a valence bond solid (VBS) state corresponding to the proton AFE state
when $J_\textrm{pro}'\gg K.$
Experimentally, an H bond will have a larger $t_\textrm{pro}$ and hence a larger $K$
than a D bond.
Therefore, an H-bonded system can exhibits a QSDL, while a D-bonded system
shows a classically ordered AFE/VBS state.
This is a generic feature of proton-driven quantum liquids and therefore may be relevant to
the experimentally observed isotope effect in $X$-Cat, where H-Cat is a QSDL while D-Cat exhibits an AFE CDW order~\cite{Isono2013,Ueda2014,Shimozawa2017qsdl}.
The observed isotope effect can be attributed to this competition.
Note that, according to the previous studies of the pyrochlore spin
liquids~\cite{Rochner2016,Savary2017}, a sufficiently large $t_\textrm{pro}$
will lead to the trivial polarized state of protons where they are simply aligned
in the direction of the ``transverse field'',
which is nothing but the proton tunneling $t_\textrm{pro}$.
In this state, the proton sector will be decoupled from the electron sector,
resulting in vanishing entanglement between them.
This state is not a quantum liquid state, although the protons are
fluctuating in the $\sigma_z$ basis. We observed reduction of the
entanglement entropy for $t_\textrm{pro}/J_\textrm{pro}\gtrsim 1$
in our numerical calculations of the finite size system (not shown).
However, the polarized proton state seems irrelevant to H-Cat since $t_\textrm{pro}$
is estimated to be 1-10 K and $t_\textrm{pro}\ll J_\textrm{pro}$
is easily satisfied~\cite{Shimozawa2017qsdl}.

Next, we consider the intertetrahedron hole hopping along the hydrogen bonds,
$t_\textrm{el}'\sum_{\langle ij\rangle}c^{\dagger}_{i\sigma}c_{j\sigma}$ as shown in Fig.~\ref{fig:lattice}(b),
within a perturbative regime $|t_\textrm{el}'|\ll |t_\textrm{el}|.$
The intertetrahedron hole hopping will induce several possible interactions between the valence
bonds on neighboring tetrahedra.
To see this,
we consider an extended Hubbard model of holes from which the $t$--$J$ model
in Eq.~\eqref{eq:H} is derived.
The Hubbard Hamiltonian for holes is
\begin{align}
\tilde{H}_\textrm{el}&=H_t+H_U,\\
H_t&=t_\textrm{el}\sum_{\boxtimes}\sum_{i,j\in\boxtimes}c^{\dagger}_{is}c_{js}
+t'_\textrm{el}\sum_{\langle ij\rangle}c^{\dagger}_{is}c_{js},\\
H_U&=U\sum_in_{i\uparrow}n_{i\downarrow}+V\sum_{\boxtimes}\sum_{i,j\in\boxtimes}n_in_j,
\end{align}
where $t'_\textrm{el}$ is an intertetrahedron hopping along a hydrogen bond $\langle ij\rangle.$
$U>0$ is an onsite interaction and $V(<U)$ is an intersite interaction within each tetrahedron.

The 2nd order perturbation in $H_t$ gives rise to several terms,
\mbox{$J_\textrm{el}[\bm{S}_i\cdot\bm{S}_j-n_in_j/4]$} with \mbox{$J_\textrm{el}=2t_\textrm{el}^2/(U-V)$},
\mbox{$J_\textrm{el}'[\bm{S}_i\cdot\bm{S}_j-n_in_j/4]$} with \mbox{$J_\textrm{el}'=2t_\textrm{el}'^2/U$},
and \mbox{$J_\textrm{el}''[n_i(n_j-1)+(n_i-1)n_j]$} with \mbox{$J_\textrm{el}''=t_\textrm{el}'^2/V$}.
The first term is an exchange interaction inside each tetrahedron, which is included in Eq.~\eqref{eq:H},
while the other two terms are new intertetrahedron interactions.
Sufficiently large $U\gg |t_\textrm{el}|$ and $V\gg |t_\textrm{el}'|$
lead to constraints on 
possible hole filling so that $n_{i}\leq 1$ for all sites $i,$ and $\sum_{i\in\boxtimes}n_i= 2$
for each tetrahedron assuming quarter filling of holes.
The resulting state can be considered as a ``tetramer Mott insulator'', as expected.

The perturbed Hamiltonian is now reduced to
\begin{align}
\tilde{H}_\textrm{el}&\simeq t_\textrm{el}\sum_{\boxtimes}\sum_{i,j\in\boxtimes}c^{\dagger}_{is}c_{js}
+J_\textrm{el}\sum_{\boxtimes}\sum_{i,j\in\boxtimes}[\bm{S}_i\cdot\bm{S}_j-n_in_j/4] \nonumber \\
&\quad +J'_\textrm{el}\sum_{\langle ij\rangle}[\bm{S}_i\cdot\bm{S}_j-n_in_j/4] \nonumber \\
&\quad +J_\textrm{el}''\sum_{\langle ij\rangle}[n_i(n_j-1)+(n_i-1)n_j].
\label{eq:H_J}
\end{align}
In the following, we only consider a singlet-pair state of holes on each tetrahedron since there is an energy gap 
$\sim J_\textrm{el}\gg J'_\textrm{el},\, J''_\textrm{el}$ between singlet and triplet states.
We note that $-(J_\textrm{el}/4)\sum_{i,j\in\boxtimes}n_in_j$ terms do not matter
in the present system
where the total fermion number inside a tetrahedron $\boxtimes$ is fixed.

The $J'_\textrm{el}$ term is nonzero only when both of the two sites
connected by the hydrogen bond $\langle i,j\rangle$ are occupied,
and similarly the $J''_\textrm{el}$ term is nonzero only when one of the two sites connected by the hydrogen bond is occupied.
Effects of the $J'_\textrm{el}$ term will be strongly suppressed by a large
interaction $g\gg J'_\textrm{el}$ between holes and protons included in the original model,
because of the energy cost due to $g$ for a configuration favored by $J'_\textrm{el}$.
On the other hand, the $J''_\textrm{el}$ term effectively adds to the interaction $g$,
because the $J''_\textrm{el}$-term also favors common local charge configurations as the $g$ term does.
Therefore, these two potential interactions between singlet pairs on the nearest-neighbor
tetrahedra
could effectively be absorbed into the interaction $g,$ or equivalently the intratetrahedron hopping $t_\textrm{el}.$
Then, it is concluded that thanks to the stability of the QSDL
for a wide range of $t_\textrm{el}/g,$
the QSDL is also stable with respect to the intertetrahedron hopping $t'_\textrm{el}.$
This behavior is numerically confirmed by exact diagonalization in the next section.

\subsection{Numerical calculations}
\begin{figure}[htb]
\includegraphics[width=8.6cm]{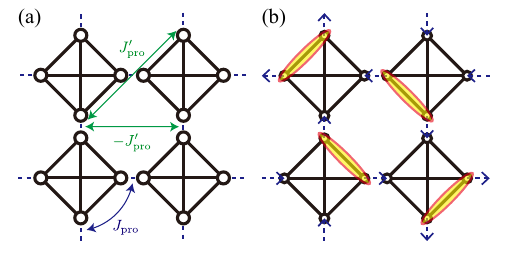}
\caption{(a) 2D lattice model used for the exact diagonalization.
Blue dashed lines represent hydrogen bonds and white circles are electron sites.
The interaction $J_\textrm{pro}$ (blue) leads to the 2 in 2 out ice rule, while an additional interaction
$J'_\textrm{pro}$ (green) favors an AFE state.
(b) Proton AFE state (blue dashed arrows) and corresponding hole VBS state (yellow ellipses) stabilized by $J'_\textrm{pro}.$}
\label{fig:lattice2d}
\end{figure}

The above discussions on the stability can be supplemented by numerical calculations
using exact diagonalization for a finite-size system with perturbations, such as the $J'_\textrm{pro}$ term.
To this end, we consider a 2D version of our model Eq.~\eqref{eq:H}
which consists of 8 proton bonds and 16 electron sites (8 holes) with a periodic boundary condition.
Although the system size is very small, this model
can reproduce much of the main results in the previous sections and thus supports the above discussions.
The Hamiltonian used in the numerical exact diagonalization is written as
\begin{align}
H_\textrm{tot}&=H+H_\textrm{pert},
\label{eq:H_ED}
\end{align}
where $H$ is a two-dimensional (2D) version of Eq.~\eqref{eq:H},
in which tetrahedra are located on vertices of a square lattice as shown in Fig.~\ref{fig:lattice2d}.
$H_\textrm{pert}$ describes perturbations,
\begin{align}
H_\textrm{pert}&=J'_\textrm{pro}\sum_{\langle ij\rangle,\langle kl\rangle}\eta^{ij}_{kl}
\sigma^z_{ij}\sigma^z_{kl}
+J'_\textrm{el}\sum_{\langle ij\rangle}[\bm{S}_i\cdot\bm{S}_j-n_in_j/4]\nonumber\\
&\quad +J_\textrm{el}''\sum_{\langle ij\rangle}[n_i(n_j-1)+(n_i-1)n_j].
\end{align}
The $J'_\textrm{pro}$ term represents an additional interaction between protons 
with properly chosen $\eta^{ij}_{kl}=\pm1$ which favors an antiferroelectric state,
and the $J'_\textrm{el},\,J''_\textrm{el}$ terms are intertetrahedron interactions between holes taken from Eq.~\eqref{eq:H_J}.
Note that for a large $g$, 
holes simultaneously exhibit the valence bond solid state out of the classical valence bond ice manifold
corresponding to the proton AFE state.
In the present finite size system, 
there are 18 ice states at $t_\textrm{pro}=J'_\textrm{pro}=J'_\textrm{el}=J''_\textrm{el}=0,$
and 16 are connected by the 2nd order perturbation in $t_\textrm{pro}.$
The 2nd order tunneling is possible because the linear system size for one direction is 2 and
the periodic boundary condition has been imposed.
Furthermore, the conservation of the total spin for each tetrahedron, $[H,\bm{S}_{\boxtimes}]=0$
for $\bm{S}_{\boxtimes}=\sum_{i\in\boxtimes}\bm{S}_i,$
greatly reduces the computational cost for the unperturbed Hamiltonian $H.$
Taking this advantage, we focus only on the subspace where
$S_{\boxtimes}=0$ for every tetrahedron also for the perturbed Hamiltonian $H_\textrm{tot},$
since there is an energy gap $\sim J_\textrm{el}$ between the $S_{\boxtimes}\equiv0$
sector and other sectors in $H.$
From now on, we omit $J_\textrm{el}$ because the triplet sectors of the original Hilbert space are all projected out.

When $J_\textrm{pro}'$ is moderately strong,
the 2D model shows an AFE state of the protons and the holes simultaneously form
a VBS state for a large $g.$
This state will have a corresponding long range order in the thermodynamic limit.
On the other hand, for a small $J_\textrm{pro}',$
it can exhibit a QSDL-like state which is a superposition of many flippable ice states.
Indeed, we find that out of the total $\mathcal{N}=18$ ice states
there are $\mathcal{N}_\textrm{flip}=16$ ice states,
which are connected by the lowest order perturbation in $t_\textrm{pro}.$
The QSDL-like state is essentially given by a superposition of these 16 reduced ice states.
The AFE/VBS state is clearly distinguished from the QSDL-like state even in the present finite-size system,
which is well-captured by the entanglement entropy $S_\textrm{EE}(t_\textrm{pro},J_\textrm{pro}')$
between protons and holes.
The entanglement entropy is given by
\begin{align}
S_\textrm{EE}&=-\textrm{Tr}_\textrm{el}\,\rho_\textrm{el}\log\rho_\textrm{el}\nonumber\\
&=-\sum_n\lambda_n\log \lambda_n,
\end{align}
where $\{\lambda_n\}$ are the squared singular values of the Schmidt decomposition of the
ground state wavefunction $\ket{\Psi}$ into the proton and electron subspaces,
\begin{align}
\ket{\Psi}=\sum_n \sqrt{\lambda_n}\ket{\textrm{pro}_n}\otimes\ket{\textrm{el}_n}.
\end{align}
The results are shown in Fig.~\ref{fig:EE}(b) where the two ground states,
AFE/VBS and QSDL states, are well characterized by $S_\textrm{EE}$.
The saturated value $S_\textrm{EE}\simeq \log 16$ for a large $t_\textrm{pro}$ means that the ground state is 
essentially given by
the superposition of 16 flippable ice states with a nearly equal weight, 
and the ground state can be considered as a QSDL in this sense.
Note that the QSDL is a stable phase extended in the parameter space, although it is parametrically small
as expected from other quantum ice systems.
In the limit of $t_\textrm{pro}\rightarrow 0$, the ground state is given by
a superposition of simple AFE/VBS states with
fourfold degeneracy due to the translational and rotational symmetries.
Therefore, $S_\textrm{EE}(t_\textrm{pro}\rightarrow0)\rightarrow \log 4$ for such a state, while
$S_\textrm{EE}\rightarrow \log \mathcal{N}_\textrm{flip}=\log 16$ deeply inside the QSDL region
with a moderately large $t_\textrm{pro}.$
This qualitatively reproduces the phase diagram in Fig.~\ref{fig:EE}(a) proposed to describe the difference
between H-Cat and D-Cat.

In the 3D thermodynamic system,
$\mathcal{N}_\textrm{flip},\, \mathcal{N}\sim e^{N_\textrm{tot}}$ and the present
QSDL-like state will be replaced by a true quantum liquid state where a macroscopic number of ice states
are involved, resulting in a large entanglement entropy, $S_\textrm{EE}\sim N_\textrm{tot}.$
Note that similar crossover behaviors in the finite size system
can also be seen in correlation functions, and those
quantities will characterize the corresponding phase transition in the thermodynamic limit.
We can also discuss the effects of an intertetrahedron hopping $t_\textrm{el}',$ and numerical results suggest
that induced intertetrahedron
valence bond interactions have only small effects on QSDL if $|t_\textrm{el}'|\ll |t_\textrm{el}|,$ and
QSDL is parametrically stable.
Detailed studies on the stability of QSDL would require more elaborate investigations, such as
cluster-based calculations~\cite{Benton2018}, and are left for a future study.

\begin{figure}[htb]
\includegraphics[width=0.95\linewidth]{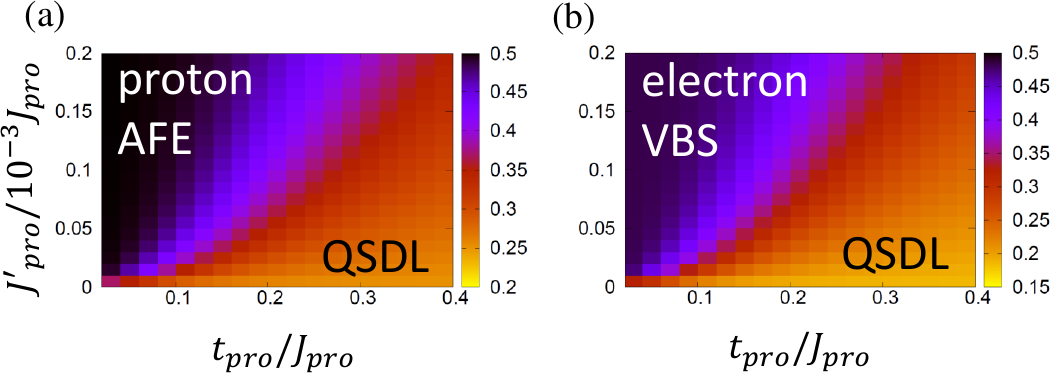}
\caption{Correlation functions (a) $\langle O_\textrm{AFE}^2\rangle$ for protons
and (b) $\langle O_\textrm{VBS}^2\rangle$ for holes.
The parameters are same as used in Fig.~\ref{fig:EE}, 
$(t_\textrm{el},g,t'_\textrm{el})=(0.3,2.0,0.0)$
in units of $J_\textrm{pro}=1.0$.}
\label{fig:LRO}
\end{figure}

Similar characterization is also possible by correlation functions corresponding to 
the long-range order of the AFE and VBS.
We consider the following two order parameters,
\begin{align}
O_\textrm{AFE}&=\frac{1}{N_\textrm{pro}}\sum_{\langle ij\rangle}\eta_{ij}\sigma^z_{ij},\\
O_\textrm{VBS}&=\frac{4}{N_\textrm{el}}\sum_{\boxtimes}\ket{\textrm{VBS}_{\boxtimes}}\bra{\textrm{VBS}_{\boxtimes}},
\end{align}
where $\eta_{ij}=\pm1$ characterizes the proton AFE configuration and
$\ket{\textrm{VBS}_{\boxtimes}}$ is the VBS configuration of the holes on a tetrahedron corresponding
to the proton AFE shown in~\ref{fig:lattice2d}(b).
The calculation results are shown in Fig.~\ref{fig:LRO}.
Note that the AFE correlation function can be evaluated explicitly for an extreme AFE state of protons,
$\ket{\textrm{AFE}}=(1/\sqrt{4})\sum_{k=1}^4\otimes_{\langle ij\rangle}\ket{\sigma^z_{ij}=\eta_{ij}^{(k)}}$ with
$\{\eta_{ij}^{(k)}\}_{k=1}^4$ being 4 degenerate AFE configurations,
and it is given by $\bra{\textrm{AFE}}O_\textrm{AFE}^2\ket{\textrm{AFE}}=1/2.$
Similarly, for an extreme QSDL state $\ket{\textrm{QSDL}}=(1/\sqrt{\mathcal{N}_\textrm{flip}})\sum_{\mathcal{C}\in
\textrm{flip}}\ket{\textrm{pro}_\mathcal{C}}\otimes\ket{\textrm{el}_\mathcal{C}}$ which
is the equal-weight superposition of $\mathcal{N}_\textrm{flip}=16$ flippable ice states,
the AFE correlation is given by
$\bra{\textrm{QSDL}}O_\textrm{AFE}^2\ket{\textrm{QSDL}}=1/4.$
These behaviors are reproduced in the numerical calculations.
Note that 
$\bra{\textrm{QSDL}}O_\textrm{AFE}^2\ket{\textrm{QSDL}}\rightarrow0$ 
in the thermodynamic limit.
The same thing happens for $\langle O^2_\textrm{VBS}\rangle.$

\begin{figure}[htb]
\includegraphics[width=0.95\linewidth]{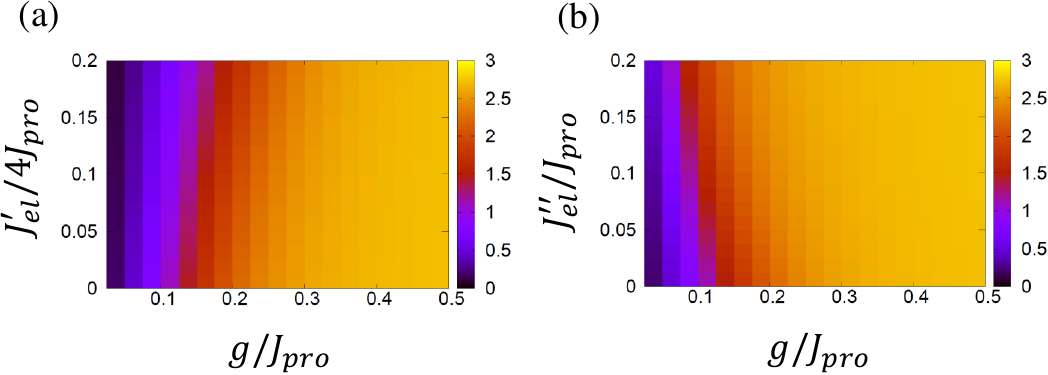}
\caption{Entanglement entropies (a) $S_\textrm{EE}(g,J'_\textrm{el})$ at $J''_\textrm{el}=0$
and (b) $S_\textrm{EE}(g,J''_\textrm{el})$ at $J'_\textrm{el}=0.$
The other parameters are $t_\textrm{pro}=0.2J_\textrm{pro},\,J'_\textrm{pro}=0.$}
\label{fig:EE2}
\end{figure}

Next, we discuss the effect of the intertetrahedron interactions $J'_\textrm{el},\,J''_\textrm{el}$ arising from
the hole hopping $|t'_\textrm{el}|\ll |t_\textrm{el}|.$
Fig.~\ref{fig:EE2} shows entanglement entropies $S_\textrm{EE}$ for different values of $(g,J'_\textrm{el})$
at $J''_\textrm{el}=0$
and $(g,J''_\textrm{el})$ at $J'_\textrm{el}=0.$
The other parameters used are $t_\textrm{pro}=0.2J_\textrm{pro},\, J'_\textrm{pro}=0.$
$S_\textrm{EE}(g,J'_\textrm{el})$ is only slightly decreased by the introduction of $J'_\textrm{el},$
because it competes with $g,$ but is negligible for a large $g.$
$S_\textrm{EE}(g,J''_\textrm{el})$ is a little enhanced by $J''_\textrm{el},$
since it favors valence bond configurations including the flippable ice states.
These behaviors are consistent with the qualitative discussion in the previous section.
Therefore, the QSDL is parametrically stable for a small intertetrahedron hole hopping $t'_\textrm{el}.$

\section{Summary and Discussions}
We have developed a theory of a hybrid quantum liquid, QSDL,
by combining the quantum proton ice and Anderson's RVB state,
motivated by the recent experimental discovery of the quantum spin liquid with proton fluctuations in H-Cat.
We proposed an idealized model and demonstrated that proton tunneling 
drives the system into 
the quantum valence bond ice,
a QSDL described by an emergent $\mathrm{U}(1)$ gauge theory.
Our theory sheds light on the essential roles of protons in the realization of QSDLs by proposing a coherent understanding
of spin-dipole coupled systems, and thus can provide a basis for future developments
not only for $X$-Cat but also for other QSDL candidate systems~\cite{Hassan2018qdl,Kitagawa2018qsl}.
For example, a hydrogen-assisted Kitaev spin liquid material H$_3$LiIr$_2$O$_6$
was recently discovered~\cite{Kitagawa2018qsl},
and proton tunneling is suggested to be important to stabilize the quantum spin liquid~\cite{Yadav2018hydrogen,Li2018hydrogen,Knolle2018random,Wang2018hydrogen}.
In addition, 
although we have been based on the idealized model Eq.(1) to elucidate the impacts of H-bonds, 
such a model itself could be designed in metal-organic frameworks (MOF). 
Indeed, there are a variety of H-bonded MOFs and nature of the H-bonds have been discussed extensively~\cite{MOF1,MOF2}. 
Realization of an ideal system for a hybrid quantum liquid by utilizing the flexibility of MOFs would be an interesting future direction.

The large entanglement predicted by our theory characterizes a QSDL in general,
and it would lead to spin-dipole coupled dynamics such as a magnetoelectric effect, which
could provide an experimental signature of the entangled spins and protons.
Other experimentally observable quantities characterizing a QSDL are dynamical ones,
which can be captured by neutron scattering or nuclear magnetic resonance.
Further investigation is necessary to discover a smoking gun signature for the strong
entanglement between the two degrees of freedom.

\begin{acknowledgments}
We thank M.~P.~A.~Fisher, Y.~Fuji, H.~Katsura, G.~Misguich, M.~Naka, M.~Oshikawa, H.~C.~Po, F.~Pollmann, T.~Senthil,
H.~Seo, N.~Shannon, M.~Shimozawa, T.~Tsumuraya, S.~Tsuneyuki, and M.~Udagawa for valuable discussions.
M.G.Y. is supported by the Materials Education program for the future leaders in Research, Industry, and Technology (MERIT), and by JSPS. 
This work was supported by JST CREST Grant Number JPMJCR19T5, Japan,
JSPS KAKENHI Grant No. JP17J05736,
No. JP17K14333, KAKENHI on Innovative Areas ``J-Physics''
[No. JP18H04318], and 
in part by the National Science Foundation under Grant No. NSF PHY-1748958.
\end{acknowledgments}


\appendix

\section{Derivation of a quantum ice model}
\label{App:A}
Here we derive the effective low-energy Hamiltonian Eq.~\eqref{eq:QIM}
from the original Hamiltonian Eq.~\eqref{eq:H}, based on the
degenerate perturbation theory with respect to $t_\textrm{pro}.$
The 6th order perturbation in $t_\textrm{pro}$ gives
\begin{align}
H_\textrm{eff}=\mathcal{P}V_t\frac{1-\mathcal{P}}{E_0-H_0}V_t \cdots V_t
\frac{1-\mathcal{P}}{E_0-H_0}V_t\mathcal{P},
\end{align}
where we decompose the original Hamiltonian $H(t_\textrm{pro})$
given by Eq.~\eqref{eq:H}
into $H(t_\textrm{pro}) = H_0 + V_t,$ and $H_0 = H(t_\textrm{pro}=0).$
$\mathcal{P}$ is a projection operator onto the degenerate ice state manifold with
an energy $E_0.$

\begin{figure}[htb]
\includegraphics[width=0.55\linewidth]{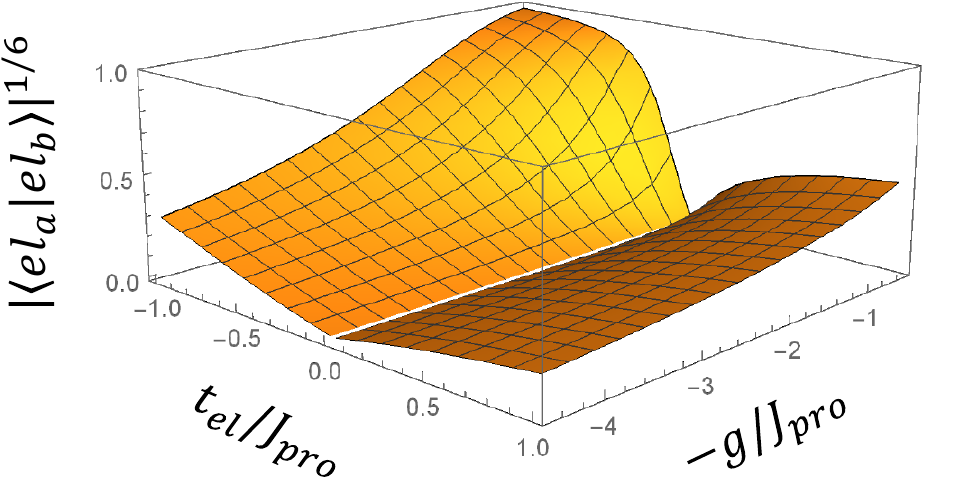}
\caption{Overlap matrix element $|\braket{\textrm{el}_a |\textrm{el}_b} |^{1/6}$
as a function of $t_\textrm{el}$ and $g$.
For large values of $g$, the overlap is linear in $t_\textrm{el}$.
}
\label{fig:ovlp}
\end{figure}

Though an intratetrahedron hole hopping $t_\textrm{el}$ has been included in $H_0$
and the perturbation itself is done for a small $t_\textrm{pro}/J_\textrm{pro},$
we concentrate on a principal contribution coming from the lowest order of $t_\textrm{el}/g.$
To do this, we take a basis set
consisting of product
states of hole and proton sectors $\{\ket{\textrm{pro}}\otimes \ket{\textrm{el}}\}.$
Each hole state $\ket{\textrm{el}}$ can be explicitly obtained by diagonalizing the hole Hamiltonian of each tetrahedron for a given
proton configuration $\ket{\textrm{pro}}.$
By inserting an identity operator, $\openone = \sum \ket{\textrm{pro}}\otimes \ket{\textrm{el}}
\bra{\textrm{pro}}\otimes \bra{\textrm{el}},$
we find for two flippable ice states $\ket{a},\,\ket{b},$
\begin{align}
h_{ab} 
&\simeq 
\frac{-12t_\textrm{pro}^6(21J^3+118GJ^2+214G^2J+120G^3)}{(J+G)^3(J+2G)^2(J+3G)(J+4G)(J+5G)} \nonumber \\
&\quad \times \braket{\textrm{el}_a | \textrm{el}_b}        
\end{align}
where $h_{ab}=\braket{a | H_\textrm{eff} | b},$ $J=4J_\textrm{pro},$ and $G=2g.$
$\ket{\textrm{el}_{a(b)}}$ is the hole wavefunction in the ice state $a(b).$
The overlap matrix element can be calculated numerically and is $O(t_\textrm{el}^6/g^6)$ for a small $|t_\textrm{el}|$
as shown in Fig.~\ref{fig:ovlp}.
Amplitude of the overlap takes the same value for any pair of electron states connected by the 6th order perturbation
    of $t_\textrm{pro},$ 
and therefore we simply denote $K\equiv -h_{ab}$ to obtain Eq.~\eqref{eq:QIM} in the main text
as $H_\textrm{eff}=-K\sum \ket{a}\bra{b}.$
The effective Hamiltonian can explicitly be expressed
in terms of hole operators $c_{is}^{(\dagger)},$
and
$\ket{\textrm{el}_{a}}\bra{\textrm{el}_{b}}
=\prod_\textrm{plaquette}c_{is}^{\dagger}c_{js} \quad (i,\,j\in \boxtimes)$ is simply a ring-hopping operator
for each hexagonal plaquette loop
in the leading order of the perturbation in $t_\textrm{el}/g.$

\section{Entanglement entropy for the Rokhsar-Kivelson state}
\label{App:B}
The entanglement entropy between protons and electrons can be calculated explicitly
for an extreme limit of a quantum spin-dipole liquid (QSDL), the Rokhsar-Kivelson (RK) state.
The RK state is the exact ground state of the ice model Eq.~\eqref{eq:QIM} in the main text with an additional potential
$\mu \sum [\ket{\circlearrowleft}\bra{\circlearrowleft } +\ket{\circlearrowright}\bra{\circlearrowright }]$
fine-tuned to $\mu=K.$
The RK state is a superposition of all possible ice states,
$\ket{\textrm{RK}}=(1/\sqrt{\mathcal{N}})\sum_\mathcal{C}
\ket{\textrm{pro}_\mathcal{C}}\otimes\ket{\textrm{el}_\mathcal{C}},$
where $\mathcal{C}$ represents an ice configuration and $\mathcal{N}$ is the dimension of the ice manifold.
$\ket{\textrm{pro}_\mathcal{C}}$ and $\ket{\textrm{el}_\mathcal{C}}$ are proton and electron wavefunctions
for a fixed ice configuration $\mathcal{C}.$
The entanglement entropy between protons and holes
is given by $S_\textrm{EE}=\log\mathcal{N}$ 
and it obeys a clear volume law $S_\textrm{EE}=O(N_\textrm{tot}).$

\bibliography{paper}

\end{document}